\documentclass{sigcomm-alternate}

\usepackage{cite,amsmath,graphicx,epsfig,subfigure,times,latexsym,algorithm2e,verbatim}

\begin{document}

\title{Dynamic Network Probes: A Stepping Stone \\to Omni Network Visibility}

\numberofauthors{2} 
\author{
\alignauthor Haoyu Song, Jun Gong, Hongfei Chen\\
      \affaddr{Huawei Technologies}\\
      \affaddr{Santa Clara, USA \& Beijing, China}\\
      \email{\{haoyu.song, gongjun, chenhongfei\}@huawei.com}
\alignauthor Tom Tofigh\\
	\affaddr{AT\&T}\\
	\affaddr{Menlo Park, USA}\\
	\email{mt3682@att.com}
}	

\maketitle
\begin{abstract}

Effective SDN control relies on the network data collecting capability as well as the quality and timeliness of the data. As open programmable data plane is becoming a reality, we further enhance it with the support of runtime interactive programming in order to cope with application dynamics, optimize  data plane resource allocation, and reduce control-plane processing pressure. Based on the latest technologies, we propose the Dynamic Network Probes (DNP) as a means to support real-time and on-demand network visibility. DNPs serve as an important building block of an integrated networking data analytics platform which involves the network data plane as an active component for in-network computing. In this paper, we show the types of DNPs and their role in the big picture. We have implemented an NP-based hardware prototype to demonstrate the feasibility and efficiency of DNPs. We lay out the research challenges and our future work to realize the omni network visibility based on DNPs.   

\end{abstract}


\section{Introduction}\label{intro}

Many of network service provider's pain points can be traced back to a fundamental issue:  the lack of network visibility. For example, the network congestion collapse can be avoided in many cases if we know exactly when and where the congestion is happening or even better, if we can precisely predict it well before any impact is made; sophisticated network attacks can be prevented through stateful and distributed network behavior analysis; in order to monetize the traffic and provide application-centric service, user flows and their interaction with networks need to be tracked and understood. 

All these pose vast needs on generalized network data, either passing through networks or generated by network devices. Not surprisingly, people start to model the network visibility as a big data problem. Traditional algorithms can still apply, but more advanced big data analytics such as machine learning can provide unlimited opportunities to mine values from network data. If we can retrieve any data of interest in real time with direct helps from the data source (i.e., the network data plane), then most problems can be solved by the comprehensive data analytics at the application plane. Therefore, our value proposition is to \emph{build a unified and general-purpose network data analytics platform with integrated data plane support to provide the omni network visibility}. This is in contrast to the ad-hoc solutions which only deal with one single problem a time with a special means to acquire the relevant data.

SDN appears to be an ideal architecture to support the omni network visibility. The logically centralized controller is at the unique vantage point to see any data in networks. However, so far the SDN controller has limited view of network data and states, because the data plane is incapable of providing enough data to sustain all application requirements. Therefore, we need to make the data plane programmable so any data can be collected if needed. More importantly, the programmability must be open to the upper layer applications so service provider can directly take advantage of this flexibility for application layer data collection.    
    
We further argue that the data plane should also support on-demand and real-time programming to meet the dynamic data needs from various runtime applications (detailed in Sec.~\ref{motiv}). This interactive programming capability is in contrast to the conventional programming with static languages. With such a data plane, we can add network probes anywhere and anytime through a standard interface. These passive probes will not alter the forwarding behavior but add the monitor points which are responsible to collect and preprocess data for applications. 

We make two major contributions in this paper. First, we devise the Dynamic Network Probe  (DNP) as a flexible and dynamic means for SDN data plane data collection. Second, we show the possibility to build a universal network data analytics platform in which network devices play an integrated role. The first contribution forms the foundation for the second one.


\section{Why Dynamic Data Collection}\label{motiv}

Data collected from data plane provide input to the SDN control loop. What data to collect is determined by the purpose of the data-consuming application. The most basic application is routing and forwarding decision for which the network topology and link states need to be collected. Other applications, such as traffic engineering, network security, network health monitoring, trouble shooting, and fault diagnosis, require different type of data. The data are either normal traffic packets that are filtered, sampled, or digested, or metadata generated by network devices to convey network states and status. In either case, data collection is meant to be passive and should not change the network forwarding behavior. 

The SDN controller analyzes the collected data and then makes decisions accordingly to actively change the network behavior. Forwarding Information Base (FIB) and Access Control List (ACL) table updates are the most notable examples, as well as Traffic Manager (TM) parameters. A few other possible changes are less obvious. For example, some applications keep refining the ways to collect data based on previous observations. The control loop algorithm continuously interacts with the data plane and modify the data source and content. We can also imagine that in some other applications, new network and packet processing functions can be enabled to alter the forwarding behavior at runtime. 

This paper concerns with the passive data collection only. Although the technology discussed in this paper can also be applied to actively modify the network behavior, we leave that topic to future work. Ideally, we want to gain the full visibility to know any states anytime anywhere in the entire network data plane. In reality, this is extremely difficult if not impossible. Theoretically, any network state can be inferred if all the traffic through it can be seen, so a simple option is to mirror all the raw traffic to servers where data analytical engine is running. However, this brute-force method requires to double the device port count and the traffic bandwidth, and poses enormous computing and storage cost. As a tradeoff, Test Access Port (TAP) or Switch Port Analyzer (SPAN) is used to selectively mirror only a portion of the overall traffic. Network Packet Broker (NPB) is deployed along with TAP or SPAN to process and distribute the raw data to various data analytical tools. There are some other ad-hoc solutions (e.g., sflow~\cite{sflow} and everflow~\cite{everflow}) which can provide sampled and digested packet data and some traffic statistics. Meanwhile, network devices also generate various log files to record miscellaneous events in the system.

When aggregating all these solutions together, we can gain a relatively comprehensive view of the network. However, the main problem here is the lack of a unified platform to deal with the general data collection problem. Moreover, each ad-hoc solution inevitably loses information due to data plane resource limitation which makes the data analytical results suboptimal, so does the follow-up data plane control based on the results. 

We also note that application's requests on data are often dynamic and realtime. When third party applications run on top of the network operator's controller, their data needs are diversified and unpredictable. Even a single application may need to constantly adjust what data to collect (e.g., an elephant flow detector continues to narrow down the flow granularities and gather their statistics). Trying to design an omnipotent system to support all possible runtime data requests is inviable because the resources required are prohibitive (e.g., even a simple counter per flow is impossible in practice). An alternative is to reprogram or reconfigure the data plane device whenever an unsupported data request appears. This is possible thanks to the recently available programmable chips~\cite{stanfordTI} and the trend to open the programmability to service providers~\cite{p4}. Unfortunately, the static programming approach cannot meet the realtime requirements due to the latency incurred by the programming and compiling process. The reprogramming process also risks breaking the normal operation of network devices.        

Then a viable solution left to us is: whenever applications request data which is unavailable in the data plane, the data plane can be configured in real time to return the requested data. That is, we do not attempt to make the network data plane provide all data all the time. Instead, we only need to ensure that \emph{any application can acquire any necessary data instantly whenever it actually asks for it.} This data-on-demand model can support effectively ``omni'' network visibility,  but it is still unthinkable with the current stiff and black-boxed data plane.  So we first introduce the recent technology advance to ground the feasibility of our vision in Sec.~\ref{tech} and \ref{pof}. 

\section{Open Programmable Data Plane}\label{tech}

\begin{figure}[!ht]
\centering
\includegraphics[width=0.7\columnwidth]{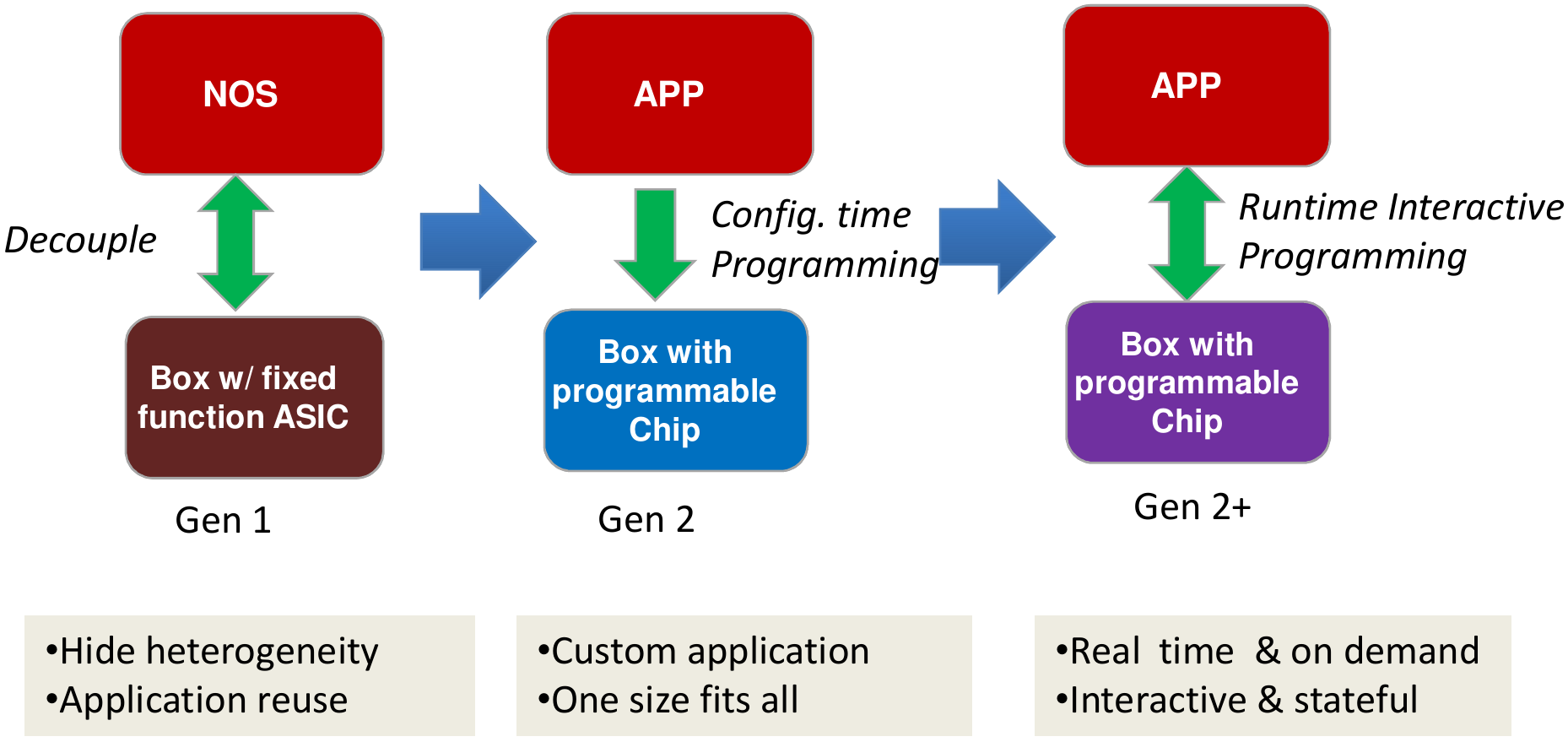}
\caption{Towards Open Programmable Data Plane}
\label{fig_opdp}
\end{figure}

Driven by SDN, network data plane is evolving to become open programmable~\cite{pof-1}. 
This means the network operators are in control of customizing the network device's function and forwarding behavior. Several ongoing trends in industry, as shown in Figure~\ref{fig_opdp},  are validating this idea. These trends are shaping new forms of network devices and inspiring innovative ways to use them.

The first trend is led by the OCP networking project~\cite{ocp}, which advocates the decoupling of the network operating system and the network device hardware. A common Switch Abstract Interface (SAI) allows applications to run on heterogeneous substrate devices. However, such devices are built with fixed function ASICs, which provide limited flexibility for application customization.  

The second trend is built upon the first one yet makes a big leap. Chip and device vendors are working on opening the programmability of the NPU, CPU, and FPGA-based network devices to network operators. Most recently, programmable ASICs has been proved feasible~\cite{stanfordTI}. High level language such as P4~\cite{p4} is developed to make the network device programming easy and fast. Now a network device can be incarnated into different functioning boxes depending on the program installed.

However, such programming process is considered static. Even a minor modification to the existing application requires to recompile the updated source code and reinstall the application. This incurs long deployment latency and may also temporarily break the normal data plane operation. Hence, although this generation of trend is still in its infancy, we have seen some of its limitations.  

Open programmable data plane should be stretched further to support runtime interactive programming in order to extend its scope of usability. Dynamic application requirements cannot be foreseen at design time, and runtime data plane modifications are required to be done in real time (for agile control loop) and on demand (to meet data plane resource constraints). Meanwhile, the data plane devices are capable of doing more complex things such as stateful processing without always resorting to controller for state tracking. This allows network devices to offload a significant portion of the data processing task and only hand off the preprocessed data to the data-requesting applications. 

\begin{figure}[!ht]
\centering
\includegraphics[width=0.4\columnwidth]{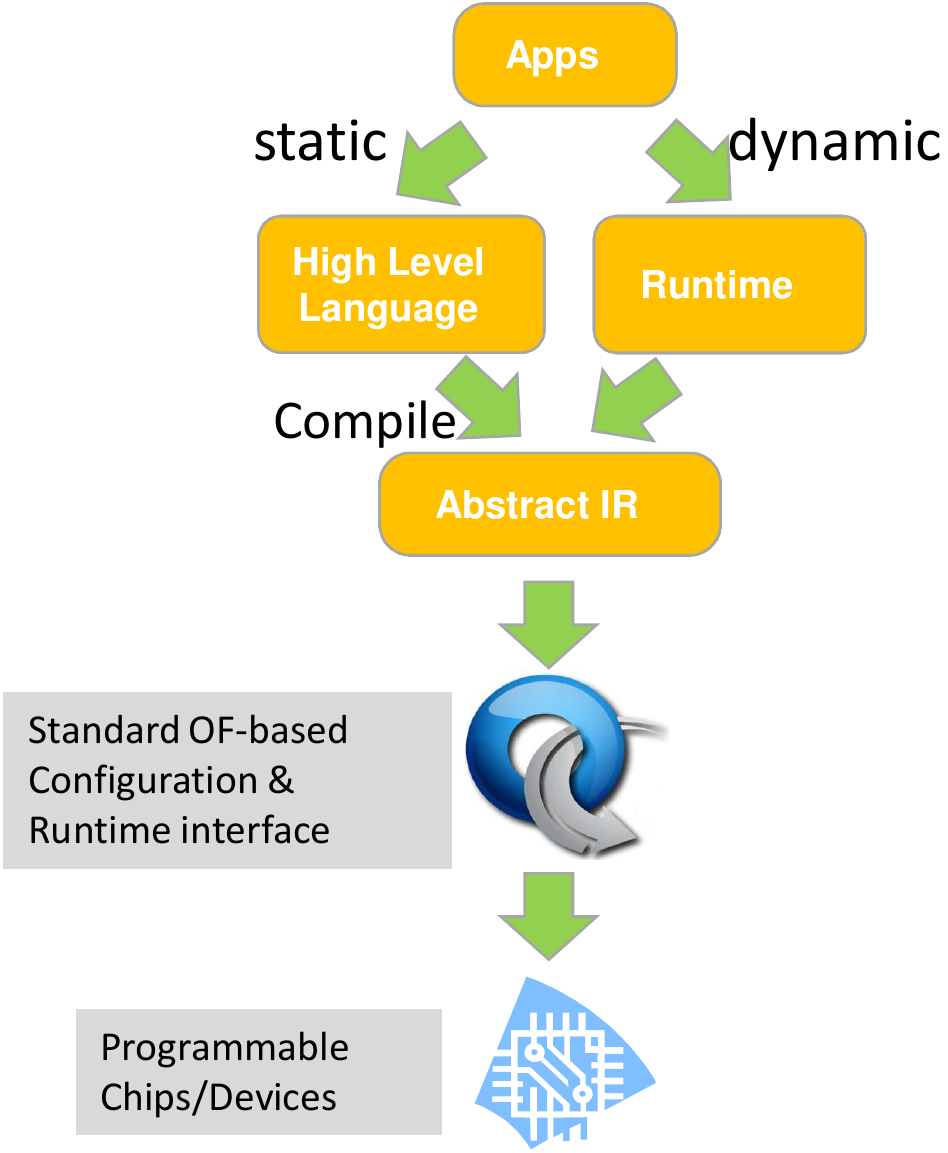}
\caption{Programming Data Plane Devices}
\label{fig_pof1}
\end{figure}  

We introduced the POF programming model to support this trend as shown in Figure~\ref{fig_pof1}~\cite{pof-2}. We still use static programing with high level languages to define the main data plane processing and forwarding function. But at runtime, whenever an application requires to make some modification to the data plane, we deploy the incremental modification directly through the runtime control channel. The key to make this dynamic and interactive programming work is to maintain a unified interface to devices for both configuration and runtime control, because both programming paths share the same data plane abstraction and use the same back-end adapting and mapping method. 

The data plane processing and forwarding function is abstracted as the standard Intermediate Representation (IR). IR contains high-level and abstract objectives that structurize the data plane components and flow logic~\cite{ir}. These objectives are not deviated too much from the high level language such as P4~\cite{p4}. This means for incremental changes, it is possible to provide APIs directly at IR level. User can therefore interactively apply data plane changes and avoid the full source-code compiling process.     

Of course, the network devices need to be flexible enough to support the interactive programming. We have implemented an NPU-based hardware prototype to support it. The virtual network device running on CPU/GPU can certainly support it easily.  ASICs and FPGAs are also possible to support it but do need some architectural innovations which will be briefly discussed in Section~\ref{conclude}.                  

\section{Interactive Programming}\label{pof}

The forwarding application in data plane can be modeled in different ways (e.g., ForCES~\cite{forces}), but so far the most popular abstract model is the match-action table pipeline\cite{openflow, pof-1, p4}. In this model, the streaming packets coming from some physical or logical input ports enter a pipeline. At each pipeline stage, some data extracted form the packet or the metadata is used as key to search a table. The matching entry triggers some associated action. The action may lead the packet to another pipeline stage or an output port. A single network device may contain multiple such pipelines, segregated by logical ports (e.g., controller, switch fabric, or black box modules). This simple yet expressive model is sufficient to describe arbitrary data plane forwarding applications. 

We make several enhancements to the basic model in order to support the runtime interactive programming. First, the actions are no longer statically associated with flow table entries as in OpenFlow~\cite{openflow} and P4~\cite{p4}. As the chief programming point, actions contain primitive packet processing instructions and control instructions. Each custom action can be dynamically loaded into network devices. Each flow entry has a pointer field which holds a pointer to an action, and a parameter field which holds the parameters used by the associated action\footnote{\smaller{The real implementation uses another level of indirection: the flow entry pointer points to an action-ID table which contains actual pointers to actions. When multiple entries share the same action, they points to the same action ID. In case these entries need to switch to a new action, only the action-ID needs to be updated.}}. This mechanism allows users to change flow actions at runtime. It also allows each action to have different number of instructions, as long as the performance and action storage can sustain the size.  With this mechanism, users are free to download new actions to change a flow's behavior on the run. Even new pipeline stages (i.e., new tables) can be inserted into the pipeline dynamically without interrupting the data path. 

Second, some data plane resources are maintained in a dynamically shared pool. Such resources include meters, counters, timers, and global registers and tables. The rationale is multifold: (1) Network devices cannot afford to statically allocated counters or meters to all the flows and flow tables. For example, a two-million-entry IP forwarding table alone can consume 64Mb memory for counters. Since it is impossible to know a priori how to reasonably allocate these limited resources at design time, counters and meters must be shared dynamically; (2) It is easy to see if a counter or meter can be allocated to multiple flows, the aggregated flow statistics and measurement can be realized with ease. Likewise, multiple counters and meters can also be allocated to a single flow so various aspects of the flow can be counted and metered with conditional instructions; (3) A global register or table is essentially a persistent storage. When it is assigned to a flow, the flow packet can assess and update its value, therefore some stateful processing can be realized. Similarly, a register can be assigned to multiple flows to enable inter-flow information sharing, and a flow can be assigned with multiple registers to hold more stateful data. 

At last, for better stateful processing support, actions are given the privilege to write tables (i.e., insert, modify, or delete table entries). This can effectively turn a table into a state table while the search key is no longer a flow signature but a state signature. This idea was originated in \cite{pof-1} and further developed in \cite{ostate}. 

\begin{figure}[!ht]
\centering
\includegraphics[width=0.6\columnwidth]{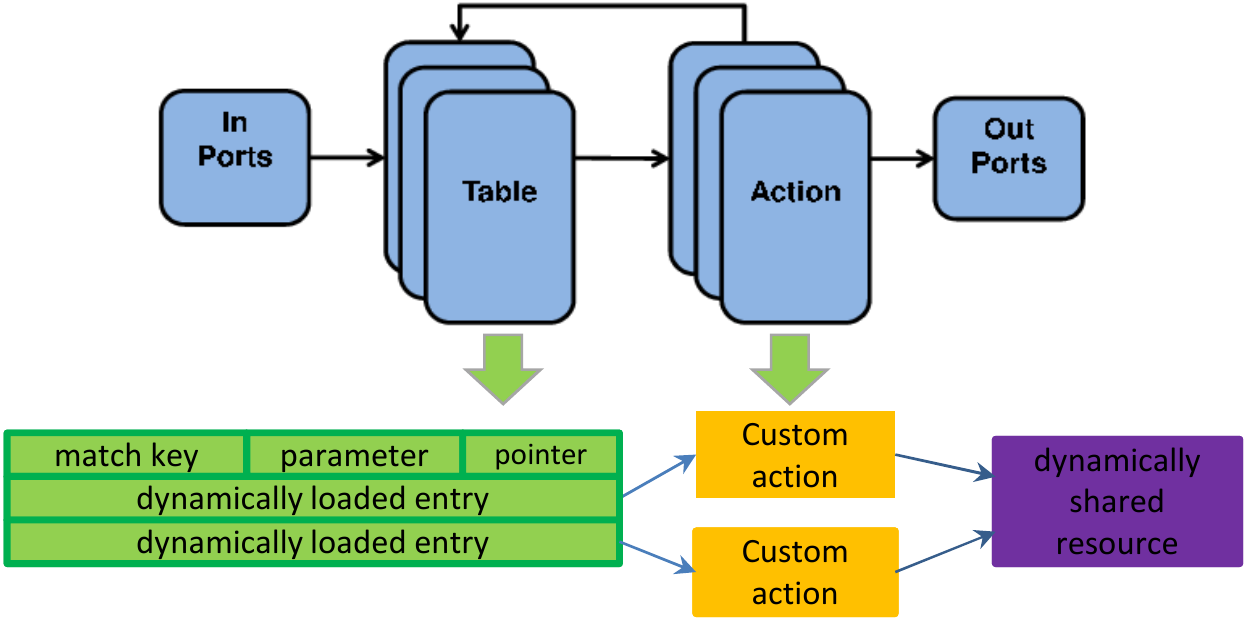}
\caption{Programming Data Plane Devices}
\label{fig_dpmodel}
\end{figure}  

Figure~\ref{fig_dpmodel} summarizes the major data plane features that support interactive programming. Such data plane is so flexible that the inattentive use of interactive programming may accidentally break the device operation or even paralyze the entire network.  However, we will see at least in one scenario we are immune from the negative effects yet can still enjoy the benefits of interactive programming. 

\section{Dynamic Network Probes}\label{probe}

Network probes are passive monitors which are installed at specific forwarding data path locations to collect specific data. DNPs are \emph{dynamically} deployed and revoked probes by applications at runtime. The customizable DNPs can collect simple statistics or conduct more complex data preprocessing. Since DNPs may require actively modifying the existing data path pipeline beyond simple flow entry manipulation,  these operations need to be done through interactive programming process. When a DNP is revoked, the involved shared resources are automatically recycled and returned back to the global resource pool.

DNPs can be deployed at various data path locations including port, queue, buffer, table, and table entry. When the data plane programmability is extended to cover other components (e.g., CPU load, fan speed, GPS coordinations, etc.), DNPs can be deployed to collect corresponding data as well. A few data plane objectives can be composed to form probes. These objectives are counter, meter, timer, timestamp, and register. Combining these with the packet filter through flow table entry configuration, one can easily monitor and catch arbitrary states on the data plane. The simplest probe is just a counter. The counter can be configured to count bytes or packets and the counting can be conditional. The more complex probes are modeled as Finite State Machines (FSM) which are configured to capture specific events\footnote{\smaller{In essence, a counter is also a simple FSM. Incrementing a counter is a stateful data plane processing.}}. We have shown how stateful processing is supported in Section~\ref{pof}. FSMs essentially preprocess the raw stream data and only report the necessary data to controller-side applications. These complex DNPs help reduce the device to controller bandwidth consumption and also alleviate the controller's processing load.  

Applications can use pull mode or push mode to access probes and collect data. The normal counter probes are often accessed via pull mode. Applications decide what time and how often the counter value is read. On the other hand, the complex FSM probes are usually accessed in push mode. When the target event is triggered,  a report is generated and pushed to the application. Below is the pseudo code of a push-mode packet counter: whenever a counter is incremented by a predefined amount, a packet is generated to report the event to the subscribing application.  

{\tiny
\begin{verbatim}
cntr[i] ++;
if(cntr[i] == threshold){
  gen_pkt(to_app, flow_id, now);
  cntr[i] = 0;
}
\end{verbatim}
}              

To install this probe, we first identify the target action. If the action $x$ exists, we generate a new action $x'$ by inserting this piece of code to the code of old action $x$. Then we download this new action $x'$ to the target device. Next we issue a command to switch the flow pointer from $x$ to $x'$. Now the probe in $x'$ takes effect and $x$ can be deleted if no longer needed.

Sometimes the target action does not exist which means a new flow entry needs to be installed along with the action. A solid example is that the application wants to gather the statistics of a new flow $f$ that does not exist. We need to choose a target table $t$ first and then analyze the normal processing action $x$ to this flow at the table $t$. We augment $x$ with the probe code to get $x'$ and download $x'$ to the target device. Next we issue commands to insert $f$ to $t$ and associate $f$ to $x'$. Note that if the flow $f$ overlaps with another flow $f'$ in table $t$, the action of $f'$ may also need to be modified.     

Timer is a special global resource. A timer can be configured to link to some action. When the time is up, the corresponding action is executed. For example, to get notification when a port load exceeds some threshold, we can set a timer with a fixed time-out interval, and link the timer to an action which reads the counter and generates the report packet if the condition is triggered. This way, the application avoids the need to keep pulling statistics from the data plane. The pseudo code of the action is as follows:

{\tiny
\begin{verbatim}
if(cntr[i] >= threshold) {
  gen_pkt(to_app, port_id, cntr[i], now);
}  
cntr[i] = 0;
\end{verbatim}
}

With the use of global registers and state tables, more complex FSM probes can be implemented. For example, to monitor the half-open TCP connections, for each SYN request, we store the flow signature to a state table. Then for each ACK packet, the state table is checked and the matched entry is removed. The state table can be periodically pulled to acquire the list of half-open connections. The application can also choose to only retrieve the counter of half-open connections. When the counter exceeds some threshold, further measure can be taken to examine if a SYN flood attack is going on. The pseudo code is shown below.

{\tiny
\begin{verbatim}
if(tcp) {
  if(syn) {
    write flow_sig to STB;
    cntr[i] ++;
  }else if(ack) {
    remove flow_sig from STB;
    cntr[i] --;  
  }  
}  
\end{verbatim}
}

Registers can be considered mini state tables which are good to track a single flow and a few state transitions. For example, to get the duration of a particular flow, when the flow is established, the state and the timestamp are recorded in a register; when the flow is teared down, the flow duration can be calculated with the old timestamp and the new timestamp. In another example, we want to monitor a queue by setting a low water mark and a high water mark for the fill level. Every time when an enqueue or a dequeue event happens, the queue depth is compared with the marks and a report packet is generated when a mark is crossed.   

Some probes are essentially packet filters which are used to filter out a portion of the traffic and mirrored the traffic to the application or some other target port for further processing. There are two ways to implement a packet filter: use a flow table that matches on the filtering criteria and specify the associated action; or directly make decision in the action. An example of the former case is to filter all packets with a particular source IP address. An example of the latter case is to filter all TCP FIN packets at edge. Although we can always use a flow table to filter traffic, sometimes it is more efficient and convenient to directly work on the action. As being programmed by the application, the filtered traffic can be further processed before being sent. Two most common processes are digest and sample, both aiming to reduce the quantity of raw data. The digest process prunes the unnecessary data from the original packet and only pack the useful information in the digest packet.  The sample process picks a subset of filtered traffic to send based on some predefined sampling criteria. The two processes can be used jointly to maximize the data reduction effect. 

An application may need to install multiple DNPs in one device or across multiple devices to finish one data analytical task. For example, to measure the latency of any link in a network. We install a DNP on the source node to generate probe packets with timestamp. We install another DNP at the sink node to capture the probe packets and report both the source timestamp and the sink timestamp to the application for link latency calculation. The probe packets are also dropped by the sink DNP. The source DNP can be configured to generate probe packets at any rate. It can also generate just one probe packet per application request. The pseudo code of the sink DNP action is as follows:

{\tiny
\begin{verbatim}
if(is_probe_packet) {
   gen_pkt(to_app, old_time, now);
   drop(this);
}  
\end{verbatim}
}

Using the similar idea, we can deploy DNPs to measure the end-to-end flow latency or trace exact flow paths. The information can be piggybacked on packets of normal traffic or on generated probe packets. Since potentially we may have many such tasks but each of such tasks may not be constantly needed and each consumes some network resources, making them dynamic is no doubt more efficient. 
In summary, DNP is a versatile tool to prepare and generate just-in-time data for data analytical applications. 

\section{Data Analytics Platform}\label{platform} 

In the past, network data analytics is considered a separate function from networks. They consume raw data extracted from networks through ad hoc protocols and interfaces. With the open programmable data plane, we expect a paradigm shift that makes the data plane be an active component of the data analytical solution. The programmable in-network computing is efficient and flexible to offload the data preprocessing through interactive data plane programming. A universal network data analytical platform built on top of this enables a tight and agile SDN control loop.  

\begin{figure}[!ht]
\centering
\includegraphics[width=0.4\columnwidth]{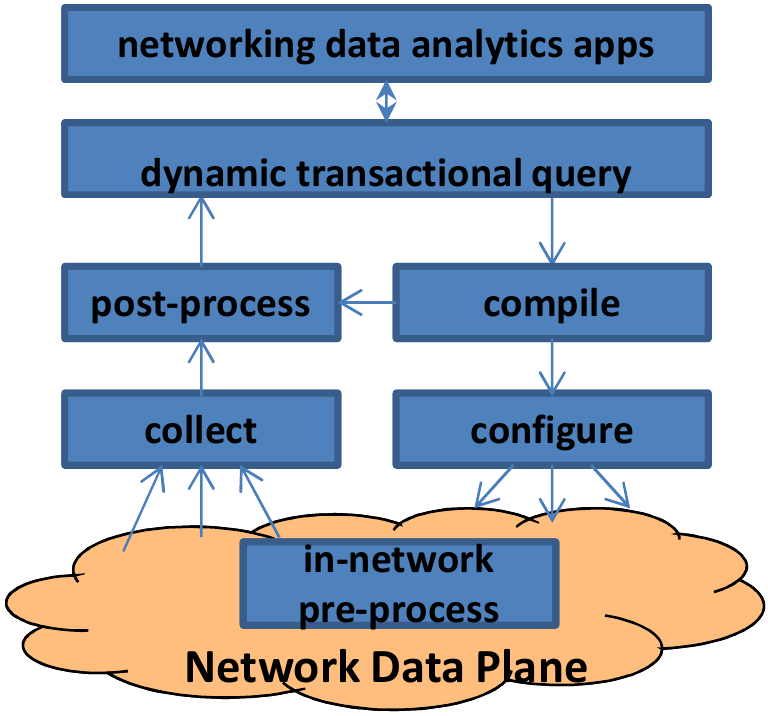}
\caption{Dynamic Networking Data Analytics}
\label{fig_platform}
\end{figure}

While DNP is a passive data plane data collection mechanism, we need to provide a declarative interface for applications to use the target-specific DNPs for data analytics. A proposed dynamic networking data analytical system is illustrated in Figure~\ref{fig_platform}. An application translates its data requirements into some dynamic transactional queries. The queries are then compiled into a set of DNPs targeting a subset of data plane devices and the instructions for data post-processing after data are collected from the data plane. After the DNPs are deployed, each DNP conducts in-network data preprocessing and feeds the preprocessed data to the collector. The collector finishes the data post-processing and presents the results to the data-requesting application.     

A query can be either continuous or one-shot. The continuous query may require the application to continuously refine the existing DNPs or deploy new DNPs. When an application revokes its queries, the idle DNP resource is released. Since one DNP may be subscribed by multiple applications, the runtime system needs to keep track of the active DNPs.    

As our future work, we aim to build a DNP-based network data analytics platform which can address multiple business requirements of network service providers, including QoE measurement, security enforcement, customer care, and network optimization. 

\section{Prototype and Evaluation}\label{eval}

We have built a hardware-based DNP prototype based on Huawei NE40E device. The device line card is equipped with a 200Gbps NPU. The POF interface protocol, essentially an extension of OpenFlow 1.4, is used as the device programming interface~\cite{pofweb}. The preliminary results are promising. We can deploy an arbitrary counter probe and start to collect results in less than $50$-ms without interrupting the normal service and impacting the forwarding performance. In contrast, to achieve the same effect with the static programming approach, we need to edit and recompile the source code of the entire design, delete the old device configuration, and download the new configuration. The compiling process consumes about $1$ second and the normal service is interrupted by about $1$ second for new program download. DNP reduces the deployment latency by 40 times. Note that this is only for a small forwarding application with 3 flow tables and 10 instruction blocks. For a larger forwarding application, the advantage of DNP is even more prominent, because deploying a DNP has a fixed cost but compiling and downloading a larger design consume more time. 

\begin{figure}[!ht]
\centering
\includegraphics[width=0.65\columnwidth]{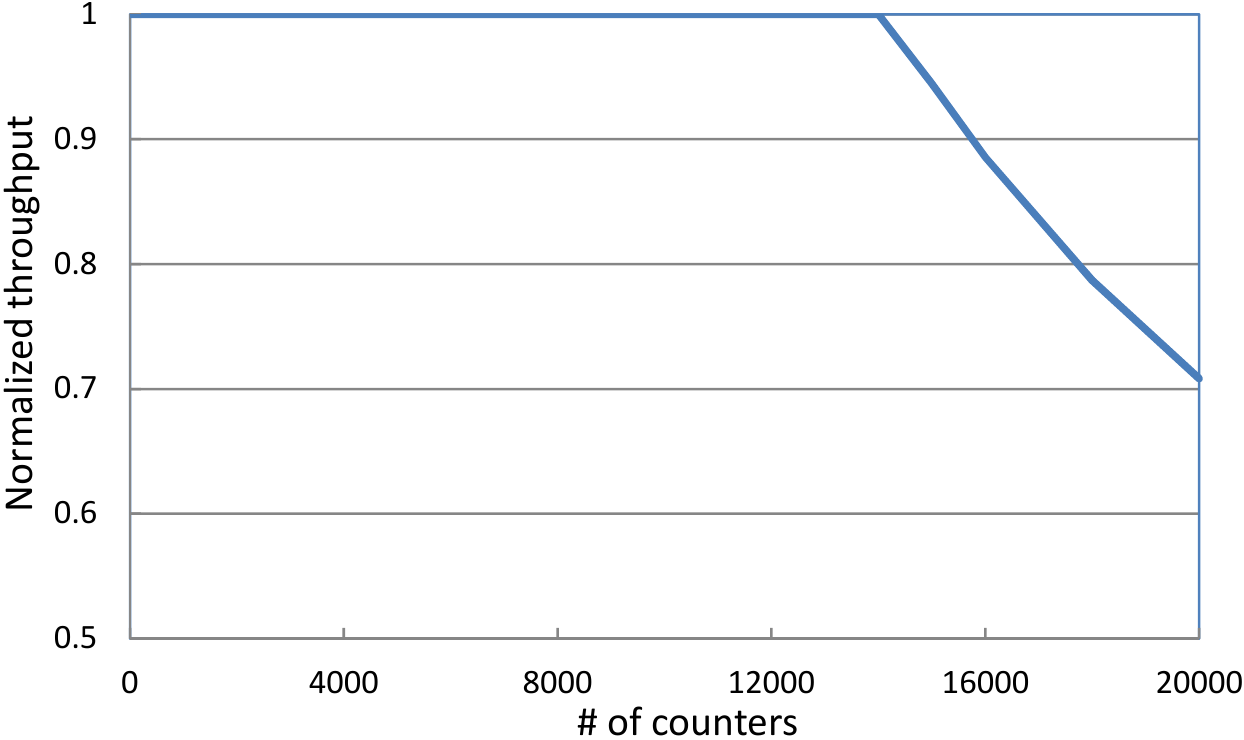}
\caption{Evaluation of DNP's Performance Impact}
\label{fig_eval}
\end{figure}

We also evaluated the DNP's performance impact to the normal forwarding throughput on our platform by keeping inserting more flow counters and observing the achievable forwarding throughput. The results are shown in Figure~\ref{fig_eval}. The factors that affect the performance include the additional instructions and memory accesses incurred by the DNPs. Our analysis shows the extra memory accesses is the dominant factor. One counter operation needs one read and one write to an SRAM block reserved for counters. Our platfom's counter memory can sustain at most 425M accesses per second. The memory bandwidth and latency interact with the limited number of cores and threads, which eventually drags down the throughput. Our evaluation shows that our platform can support line speed forwarding with up to 14K flow counters.  When more complex DNPs are applied, we expect the performance impact will be noticeable with fewer DNPs. The results also serve as an evidence why dynamic rather than static probes are needed. 

Interested readers may use the open source software provided at POF website~\cite{pofweb} to conduct the similar experiments on a software target.

\section{Challenges}\label{gap}

Many technique challenges need to be addressed to realize DNP and the universal network data analytics platform on general SDN data plane. We list a few here and also provide our initial thoughts on potential solutions. 

a) Allowing applications to modify the data plane has security and safety risks (e.g., DoS attack). The counter measure is to supply a standard and safe API to segregate applications from the runtime system and provide applications limited accessibility to the data plane. Each API can be easily compiled and mapped to standard DNPs. An SQL-like query language which adapts to the stream processing system might be feasible for the applications.

b) When multiple correlated DNPs are deployed across multiple network devices or function blocks, or when multiple applications request the same DNPs, the deployment consistency needs to be guaranteed for correctness. This requires a robust runtime compiling and management system which keeps track of the subscription to DNPs and controls the DNP execution time and order.    

c) The performance impact of DNPs must be evaluated before deployment to avoid unintentionally reducing the forwarding throughput. Fortunately, the resource consumption and performance impact of standard DNPs can be accurately profiled in advance. A device is usually over provisioned and is capable of absorbing extra functions up to a limit. Moreover, programmable data plane allows users to tailor their forwarding application to the bare bones so more resources can be reserved for probes. The runtime system needs to evaluate the resulting throughput performance before committing a DNP. If it is unacceptable, either some old DNPs need to be revoked or the new request must be denied.  
 
d) While DNP is relatively easy to be implemented in software-based platform (e.g., NPU and CPU), it is harder in ASIC-based programmable chips. Architectural and algorithmic innovations are needed to support a more flexible pipeline which allows new pipeline stage, new tables, and new custom actions to be inserted at runtime through hitless in-service updates. An architecture with shared memory and flexible processor cores might be viable to meet these requirements. Alternatively, DNPs can be implemented using an ``out-of-band'' fashion. That is, some reserved pipeline stages are dedicated for DNPs. Relevant data and/or packets are configured to be passed to these stages for processing and counting.    

\section{Related Work}\label{related}

Mainly working on fixed function network devices, sFlow~\cite{sflow}, Planck~\cite{planck}, FleXam~\cite{flexam}, and Everflow~\cite{everflow} are all packet-based mirroring and sampling techniques, through setting flow filters or relying on particular sampling policies. Such techniques can only provide partial data plane visibility with information loss. NetSight~\cite{netsight} aggressively collects the complete packet history in order to achieve full network visibility with the significant storage and computing cost at control plane. 

Some other work (e.g., TPP~\cite{minions}, INT~\cite{int, int2}, and FlowRadar~\cite{flowradar}) builds on programmable network devices, therefore non-packet data can be retrieved from data plane through programming means. TPP revives the idea of active network~\cite{actnet} but keeps it simple. It allows packets to carry a tiny program in header so the program instructions can be executed by network devices and data collected along the forwarding path. INT realizes the similar idea by using P4~\cite{p4} as the high level programming language. FlowRadar can maintain full flow statistics with succinct data structure. This group of techniques falls into the static programming category in which the custom functions are predefined at design time. To collect data that were not supported by the current design, one has to reprogram the data plane.  

Some techniques are introduced to enhance the data plane data processing capability. OpenSketch~\cite{osketch} proposes a set of data plane primitives to facilitate the network measurement. OpenState~\cite{ostate} devises a stateful data plane programming abstraction which can be used to implement FSMs. InSP~\cite{insp} provides a generic data plane packet generation API which can be used to preprocess and encapsulate collected data from data plane. DPT~\cite{dpt} suggests to augment a timestamp header to all packets for various data analytical applications.    

Interactive control on what data to collect based on network dynamics is discussed in some work (e.g., TPP~\cite{minions}, DREAM~\cite{dream}, and Mozart~\cite{mozart}). However, due to the lack of interactive programming capability, new data can only be collected through selecting existing probes or configuring new flow table entries. 

Gigascope~\cite{gigascope} and Path Query~\cite{pathquery} provide SQL-like languages for applications to initiate interactive queries for network states and data. Frenetic queries manipulate flow entries with the assumption that each flow entry has its counters and rely on the controller-side runtime system to aggreagte the data collected~\cite{frenetic}. The high-level and network-wide queries can be compiled into DNPs to collect preprocessed data direclty through data plane programming. While the network data plane is naturally a data streaming system, a Data Stream Management System (DSMS)~\cite{dsms} would rely on DNPs to realize in-network sampling and sketching techniques. 

Based on eBPF, IO-Visor~\cite{iovisor} is a kernel I/O and networking infrastructure which can be used to implement virtual switch. It supports dynamic tracing by installing virtual probes converted from high level programs at runtime. This is inline with the idea of DNP.

\section{Conclusion}\label{conclude}

DNP is enabled by the most recent technology advances: open programmable data plane and interactive data plane programming. It takes advantage of the data plane processing capability and provides real-time and on-demand network visibility at low cost and high performance. We believe DNP is a solid stepping stone to a network data analytical platform which can help service providers to gain deeper insight and mine more values from their networks. 

\small{
\bibliographystyle{abbrv}
\bibliography{sigproc}  
}
\end{document}